\documentclass[10pt,conference]{IEEEtran}

\usepackage{flushend}
\usepackage{lipsum}  
\usepackage{multirow}
\usepackage{caption}
\usepackage{subcaption}
\usepackage[dvipsnames]{xcolor}

\usepackage{cite}
\usepackage{amsmath,amssymb,amsfonts}
\usepackage{algorithmic}
\usepackage{graphicx}
\usepackage{textcomp}
\usepackage{xcolor}
\def\BibTeX{{\rm B\kern-.05em{\sc i\kern-.025em b}\kern-.08em
    T\kern-.1667em\lower.7ex\hbox{E}\kern-.125emX}}

\usepackage{url}
\usepackage{hyperref}
\hypersetup{
    colorlinks=true,
    citecolor=black,linkcolor=black,
    filecolor=black, urlcolor=cyan,
}

\usepackage{tcolorbox}
\usepackage{dblfloatfix}
\usepackage{booktabs}

\newcommand{\eg}{\emph{e.g.,}}
\newcommand{\ie}{\emph{i.e.,}}
\newcommand{\etal}{\emph{et al.}}

\newlength\MAX  
\newcommand*\Chart[1]{~\rlap{\textcolor{black!20}{\rule{\MAX}{2ex}}}\rule{#1\MAX}{2ex}}

\newtcolorbox{boxH}{
    colback = white!90!gray, 
    colframe = black, 
    boxrule = 0pt, 
    leftrule = 3pt
}

\begin{document}

\title{Exceptional Behaviors:\\How Frequently Are They Tested?}


\author{
\IEEEauthorblockN{Andre Hora}
\IEEEauthorblockA{
\textit{Department of Computer Science, UFMG}\\
Belo Horizonte, Brazil \\
andrehora@dcc.ufmg.br}
\and
\IEEEauthorblockN{Gordon Fraser}
\IEEEauthorblockA{
\emph{University of Passau}\\
Passau, Germany \\
gordon.fraser@uni-passau.de}
}

\maketitle

\begin{abstract}

Exceptions allow developers to handle error cases expected to occur infrequently.
Ideally, good test suites should test both normal and exceptional behaviors to catch more bugs and avoid regressions.
While current research analyzes exceptions that propagate to tests, it does not explore other exceptions that do not reach the tests.
In this paper, we provide an empirical study to explore how frequently exceptional behaviors are tested in real-world systems.
We consider both exceptions that propagate to tests and the ones that do not reach the tests.
For this purpose, we run an instrumented version of test suites, monitor their execution, and collect information about the exceptions raised at runtime.
We analyze the test suites of 25 Python systems, covering 5,372 executed methods, 17.9M calls, and 1.4M raised exceptions.
We find that 21.4\% of the executed methods do raise exceptions at runtime.
In methods that raise exceptions, on the median, 1 in 10 calls exercise exceptional behaviors.
Close to 80\% of the methods that raise exceptions do so infrequently, but about 20\% raise exceptions more frequently.
Finally, we provide implications for researchers and practitioners.
We suggest developing novel tools to support exercising exceptional behaviors and refactoring expensive \texttt{try/except} blocks.
We also call attention to the fact that exception-raising behaviors are not necessarily ``abnormal'' or rare.

\end{abstract}




\section{Introduction}

Exceptions are a programming construct that allows developers to handle error cases expected to occur infrequently, without cluttering the code with unnecessary if/else checks.
By using exceptions, developers can improve the program's robustness by enabling the detection, reporting, handling, and correction of exceptional behaviors~\cite{dalton2020exceptional, marcilio2021java}.
Good test suites should ideally test both normal and exceptional behaviors to catch more bugs and avoid regressions~\cite{reid1997empirical, kaner2013domain, aniche2022effective, khorikov2020unit, winters2020software, aniche2021developers}.
However, in practice, it is well-known that developers are more likely to test normal behaviors than exceptional ones~\cite{teasley1994software, salman2019controlled, causevic2013effects, edwards2014student, aniche2021developers, mohanani2018cognitive, leventhal1993positive, bijlsma2021students, garousi2018smells, bai2021students, goffi2016automatic, hora2024monitoring}, which may decay the test suite's effectiveness in finding bugs~\cite{marcilio2021java}.

\begin{figure}[t]
     \centering
     \begin{subfigure}[b]{0.48\textwidth}
         \centering
         \includegraphics[width=\textwidth]{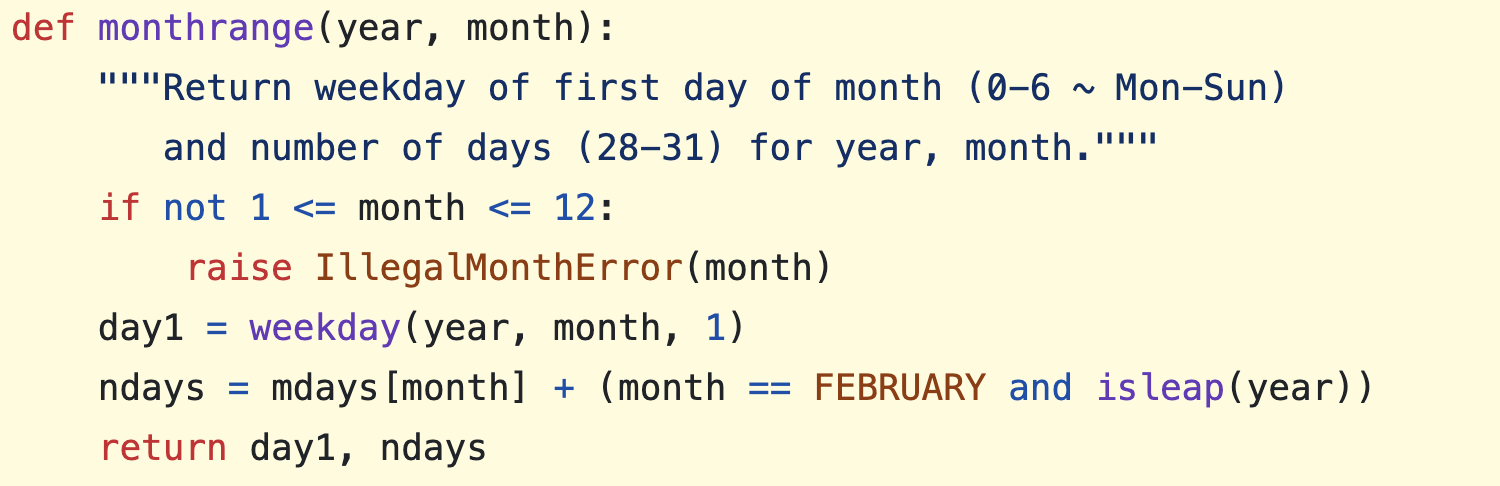}
         \caption{Method monthrange (CPython). Exception \texttt{Illegal\-Month\-Error} is rarely raised when executing the test suite.}
         \label{fig:ex1a}
     \end{subfigure}
     \par\medskip
     \begin{subfigure}[b]{0.48\textwidth}
         \centering
         \includegraphics[width=\textwidth]{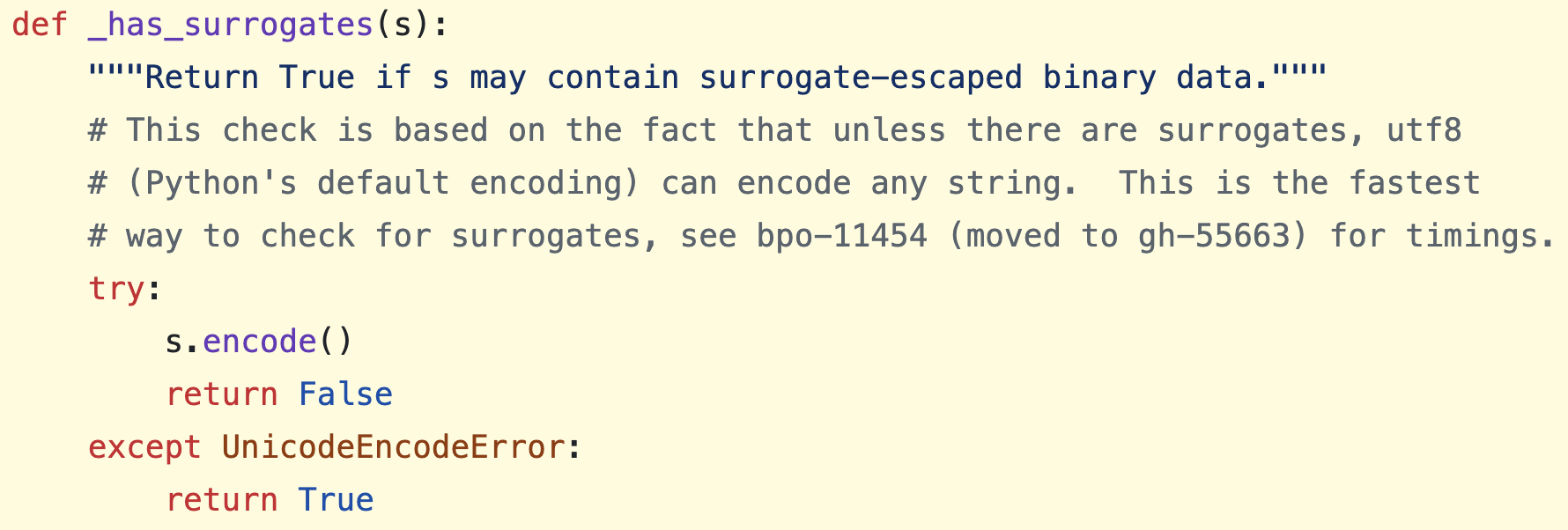}
         \caption{Method \_has\_surrogates (CPython). Exception \texttt{Unicode\-Encode\-Error} is rarely raised when executing the test suite.}
         \label{fig:ex1b}
     \end{subfigure}
     \par\medskip
     \begin{subfigure}[b]{0.48\textwidth}
         \centering
         \includegraphics[width=\textwidth]{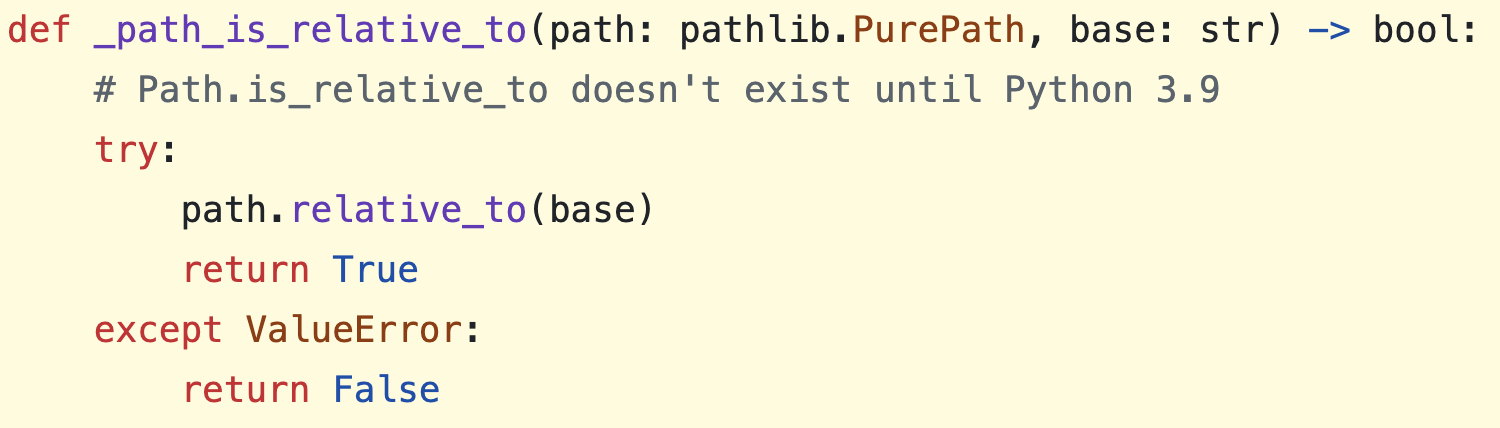}
         \caption{Method \_path\_is\_relative\_to (Flask). Exception \texttt{Value\-Error} is almost always raised when executing the test suite.}
         \label{fig:ex1c}
     \end{subfigure}
    \caption{Examples of methods with exceptional behaviors.}
    \label{fig:examples}
\end{figure}

Figure~\ref{fig:ex1a} shows method \texttt{monthrange},\footnote{\url{https://github.com/python/cpython/blob/950fab46/Lib/calendar.py\#L161}} which returns the weekday and number of days for a given \emph{year} and \emph{month}, raising an exception when \emph{month} is not between 1 and 12.
This method is called 221 times by its test suite, but only 3 calls exercise the exception.
In the case method \texttt{month\-range} is directly called by a test, it should be verified with an \texttt{assert\-Raises} to check that the exception \texttt{Illegal\-Month\-Error} gets raised.
Those tests are named ``exceptional tests'', that is, tests that expect exceptions to be raised~\cite{dalton2020exceptional, marcilio2021java}.
Method \texttt{month\-range} is exercised by three exceptional tests.\footnote{\url{https://github.com/python/cpython/blob/950fab46/Lib/test/test_calendar.py\#L894-L906}}


On the other hand, methods \texttt{\_has\_\-surrogates}\footnote{\url{https://github.com/python/cpython/blob/950fab46/Lib/email/utils.py\#L47}} and \texttt{\_path\_\-is\_\-relative\_to}\footnote{\url{https://github.com/pallets/flask/blob/2fec0b20/src/flask/sansio/scaffold.py\#L709}} presented in Figures~\ref{fig:ex1b} and~\ref{fig:ex1c} handle the raised exception with \texttt{try/except} blocks.
Thus, the raised exceptions do not propagate to the tests.
That is, from the test perspective, the exceptions \texttt{Unicode\-Encode\-Error} and \texttt{Value\-Error} are unnoticed, and we are not aware whether they get raised.
One important difference between both methods is the frequency of exception raising.
Method \texttt{\_has\_\-surrogates} is called 32,846 times by its test suite, but the exception is raised only in 0.6\% of the calls.
In contrast, method \texttt{\_path\_\-is\_\-relative\_to} raises the exception in 98\% of the calls (441 out of 448 calls).
These examples show that, at the method level, exceptions may be frequently or infrequently raised at runtime.
Moreover, when running a test suite, multiple exceptions may be raised at runtime, including the ones that are unnoticed by the tests because they are handled locally.
This means that a test suite may exercise exceptional behaviors even without explicitly asserting on the raised exceptions (\eg~using \texttt{assertRaises}).

While prior research focuses on analyzing exceptions that propagate to tests (\ie~exceptional tests~\cite{dalton2020exceptional, marcilio2021java}), it does not explore exceptions  that are \emph{not} propagated to tests.
Moreover, to our knowledge, no study has deeply explored the frequency of exception-raising at runtime from the test perspective.
A better understanding of these aspects could provide insights into how developers handle exceptional behaviors and help drive the creation of novel testing tools.

In this paper, we provide an empirical study to explore how frequently exceptional behaviors are tested in real-world systems.
We consider both exceptions that propagate to tests and the ones that do not reach the tests.
For this purpose, we run an instrumented version of test suites, monitor their execution, and collect information about the exceptions raised at runtime.
Specifically, we analyze the test suites of 25 Python systems, covering 5,372 executed methods, 17.9M calls, and 1.4M raised exceptions.
We propose three research questions to explore exceptional behaviors:

\begin{itemize}

   \item \textbf{RQ1: How many methods raise exceptions at runtime?}
    21.4\% of the executed methods do raise exceptions at runtime.
    On the median, methods that raise exceptions are called 4x more often and execute 3x more paths than those that do not raise exceptions.
   
   \item \textbf{RQ2: How frequently do calls on exception-raising methods actually lead to exceptions?}
    In methods that raise exceptions at runtime, 1 in 10 calls exercise exceptional behaviors on the median.
    Close to 80\% of the methods that raise exceptions do so infrequently, while about 20\% raise exceptions more frequently.

   \item \textbf{RQ3: How do exception-raising methods and calls vary by system?}
    Most systems (22 in 25) contain more exception-free than exception-raising methods.
    Moreover, most systems (19 in 25) have a median proportion of exception-raising calls per method below 30\%.
   
   
\end{itemize}

Based on our results, we discuss practical implications.
\underline{First}, we envision the development of novel tools to support exercising exceptional behaviors more effectively.
For example, such tools could identify the tests that cover exceptional cases, including exceptions not propagated to tests.
\underline{Second}, we reveal that a test that exercises an exception-raising method does not necessarily indicate it is testing an ``abnormal'' behavior.
For some methods, raising an exception may be part of the method's ``normal'' behavior.
Therefore, researchers working on exceptional behavior testing~\cite{dalton2020exceptional, marcilio2021java, lima2021assessing, zhang2024generating, yoshioka2023exceptional} should be aware of such methods to avoid failing to detect abnormal behaviors.
\underline{Third}, we recommend a refactoring to replace expensive \texttt{try/except} blocks.
We foresee that future research on refactoring~\cite{tsantalis2009identification, al2015identifying, alomar2024refactor, fan2023large, pomian2024together, shirafuji2023refactoring} could leverage the execution frequency of some language constructors (\eg~\texttt{try/except} blocks) to detect refactoring opportunities

\smallskip

\noindent\emph{Contributions.} The contributions of this study are twofold.
First, we propose an empirical study to explore the frequency of exception-raising at runtime from the test perspective.
Second, we provide implications for researchers and practitioners.

\section{Study Design}

\subsection{Case Studies}

In this study, we aim to study real-world software systems.
Thus, we select Python systems that are largely adopted.
We focus on Python because it is among the most popular programming languages nowadays,\footnote{\url{https://www.tiobe.com/tiobe-index}} and it has a rich software ecosystem.
Table~\ref{tab:systems} presents the 25 selected systems.
For a larger diversity of projects, we select two types of systems: (1) popular systems and (2) Python libraries.

\begin{table}[t]
\centering
\caption{Selected systems.}
\begin{tabular}{ll rr}
\toprule
\textbf{System} & \textbf{Short Description} & \textbf{Methods} & \textbf{Tests} \\\midrule
Pylint & static code analyzer       & 1,537	& 1,822 \\
Rich & rich text library            & 589	& 758 \\
Error Corrector & popular console error corrector   & 550	& 1,887 \\
BentoML & machine learning platform & 319	& 169 \\
Flask & web framework               & 284	& 478 \\
DateUtil & date and time library    & 241 & 2,029 \\
Requests & HTTP library             & 174	& 578 \\
Jupyter Client & Jupyter protocol client APIs &  138	& 87 \\
Cookiecutter & template handler     & 66	& 322 \\
Six & compatibility library         & 32	& 199 \\
\midrule
email	& email message manager     & 381 &	1,666 \\
logging		& logging facility      & 215 & 208\\
argparse & command-line interfaces  & 126 &	1,685 \\
collections	& datatype container    & 112 &	111\\
pathlib	& OO filesystem paths       & 97 & 449 \\
tarfile	& tar reading and writing   & 89	& 496 \\
configparser & configuration file parser    & 82 & 341\\
calendar	& calendar helpers      & 63 & 72\\
ftplib		& FTP protocol client   & 51 & 94\\
difflib		& diff library          & 47 & 51\\ 
imaplib		& IMAP4 protocol client & 47 & 103\\
smtplib		& SMTP protocol client  & 43 & 82\\
os			& operating system interfaces  & 41	& 316\\
gzip		& gzip reading and writing & 32	& 61\\
csv			& CSV reading and writing & 15 & 113\\ \midrule
Total       & & 5,372 & 14,177 \\
\bottomrule
\end{tabular}
\label{tab:systems}
\end{table}

\noindent\emph{Popular systems.}
The 10 selected systems presented at the top of Table~\ref{tab:systems} have thousands of GitHub stars, millions of clients, and, in some cases, billions of downloads, highlighting their relevance.\footnote{Download values are based on PePy: \url{https://pepy.tech}.}
For example,
DateUtil is a date library used by close to 1M GitHub projects. 
Pylint is the most popular code analyzer in Python, with over 400M downloads.
Rich is a text library with 40K stars and around 200M downloads.
Requests is an HTTP library with 50K stars and 6,4B downloads.
Flask is a popular web framework, with 1,3M users on GitHub and around 2B downloads.
Cookiecutter is a template handler with 20K stars.
Six is a compatibility library used by 1,5M projects with 6,5B downloads.
Lastly, Jupyter Client provides Jupyter protocol client APIs, having around 500M downloads.

\noindent\emph{Python libraries.}
We also analyze the 15 libraries presented at the bottom of Table~\ref{tab:systems}, which belong to the Python Standard Library and are hosted on the CPython repository.

\subsection{Monitoring Methods Executed by Tests}
\label{monitoring}

To extract information about exceptions raised at runtime, we need to run an instrumented version of the test suite, monitoring the method execution and collecting their raised exceptions.
For this purpose, we rely on SpotFlow~\cite{spotflow}, a tool that performs dynamic analysis in Python and collects runtime data, such as argument values, return values, and raised exceptions at the method level.
In short, this tool is implemented with the support of the standard system’s trace function \texttt{sys.settrace}~\cite{python_sys_trace}, which is the basis for performing runtime analysis in Python~\cite{msr2021_test_coverage, hora2023excluding}.
The trace function registers a hook that gets called at every executed line of code and function call, allowing the recording of method-level runtime data, such as calls and raised exceptions.

Therefore, with the support of SpotFlow, we run an instrumented version of the 25 selected test suites and detect that 5,372 application methods are executed, as detailed in Table~\ref{tab:systems}.

\subsection{Collecting Data from Executed Methods and Calls}
\label{monitoring}

While monitoring the tests, we gather data on methods and calls that raise exceptions at runtime.

\textbf{Exception-raising methods:}
For each method, we record whether it raises any exception at runtime.
Methods that do not raise exceptions are \emph{exception-free methods}, while methods that raise at least one exception are \emph{exception-raising methods}.

\textbf{Exception-raising calls:}
For each method call, we record whether it results in an exception being raised, referring to such calls as \emph{exception-raising calls}.
For each method, we compute both the absolute and relative number of exception-raising calls.
The relative number is the ratio of exception-raising calls to the total number of calls.

In total, the 5,372 executed methods receive 17.9M calls and raise 1.4M exceptions at runtime.
On the median, each executed method receives 33 calls.
This data is further explored in our research questions.
Our dataset is publicly available at: \url{https://doi.org/10.5281/zenodo.14187323}.

\subsection{Research Questions}

We propose three research questions to explore exception-raising frequency at runtime.
In RQ1, we analyze how many methods raise exceptions.
In RQ2, we assess how frequently calls lead to exceptions.
RQ3 explores the variation of exception-raising methods and calls per system.
\textbf{Rationale:}
Each research question addresses the frequency of the raised exceptions at a distinct granularity level.
RQ1 focuses on the method level, while RQ2 on the call level.
RQ3 analyzes methods and calls but on the system level.
A better understanding of these aspects could provide insights into how developers handle exceptional behaviors, potentially guiding the development of novel testing tools.

\section{Results}

\subsection{RQ1: How many methods raise exceptions at runtime?}

Considering the 25 selected Python systems, 5,372 application methods are directly or indirectly executed by their test suites.
Among these methods, 21.4\% (1,150) do raise exceptions at runtime, while 78.6\% (4,222) do not raise any exceptions, as detailed in Table~\ref{tab:rq1}.

\begin{table}[t]
\centering
\caption{Summary of methods raising exceptions at runtime.}
\begin{tabular}{l rrl}
\toprule
\textbf{Categories} & \textbf{\#} & \textbf{\%} & \\ \midrule
Exception-free methods & 4,222 & 78.6\% & \Chart{0.786} \\
Exception-raising methods & 1,150 & 21.4\% & \Chart{0.214} \\ \midrule
All & 5,372 & 100\% & \Chart{1.0} \\
\bottomrule
\end{tabular}
\label{tab:rq1}
\end{table}

Table~\ref{tab:exe_types} presents the distribution of the exception types of raised exceptions.
In total, we find that 200 distinct exception types are raised at runtime.
Most exceptions are raised by a single method (73 in 200).
Moreover, 62 exceptions are raised by 2--3 methods, 40 exceptions by 4--9 methods, and only 25 exceptions by 10 or more methods.
The most raised exception types are the generic ones: \texttt{Value\-Error}, \texttt{Generator\-Exit}, \texttt{Type\-Error}, \texttt{Key\-Error}, and \texttt{Stop\-Iteration}.
On the other hand, exception types raised by single methods are very specific, such as \texttt{SMTP\-Sender\-Refused}, \texttt{Invalid\-JSON\-Error}, \texttt{Empty\-Header\-Error}, \texttt{Illegal\-Weekday\-Error}, and \texttt{NoEmoji}.

\begin{table}[t]
\centering
\caption{Distribution of exception types.}
\begin{tabular}{l c cccc}
\toprule
& \multirow{2}{*}{\textbf{Total}} & \multicolumn{4}{c}{\textbf{Methods}} \\ \cmidrule{3-6}
& & 1 & 2--3 & 4--9 & 10+ \\ \midrule
\textbf{Types of raised exceptions} & 200 & 73 & 62 & 40 & 25 \\
\bottomrule
\end{tabular}
\label{tab:exe_types}
\end{table}

\begin{figure}[t]
    \centering
    \includegraphics[width=0.45\textwidth]{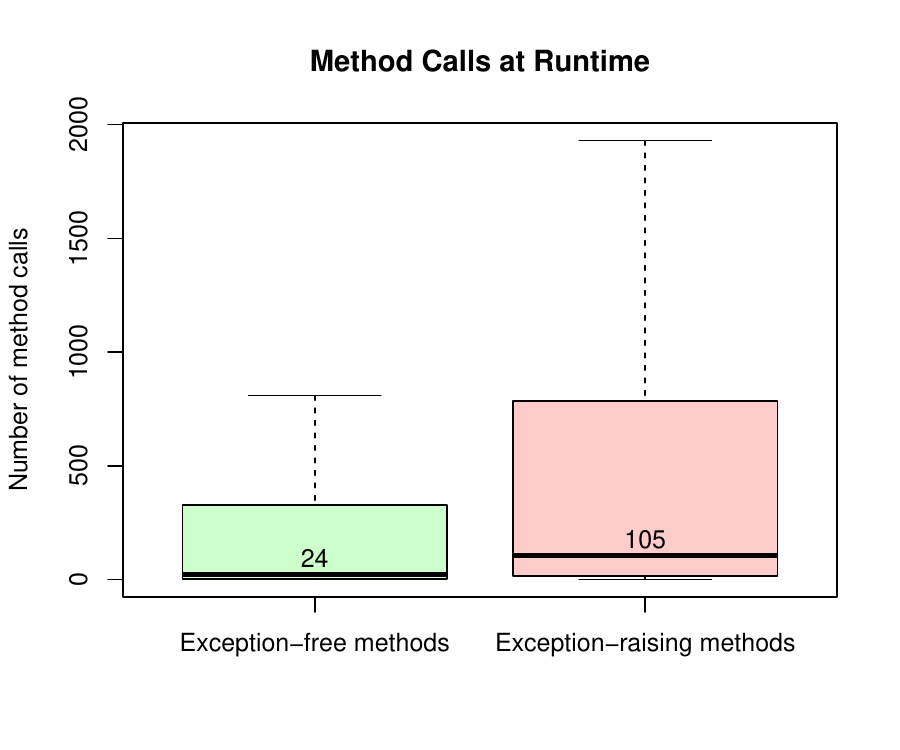}
    \caption{Distribution of method calls at runtime.}
    \label{fig:box_calls}
\end{figure}

\begin{figure}[t]
    \centering
    \includegraphics[width=0.45\textwidth]{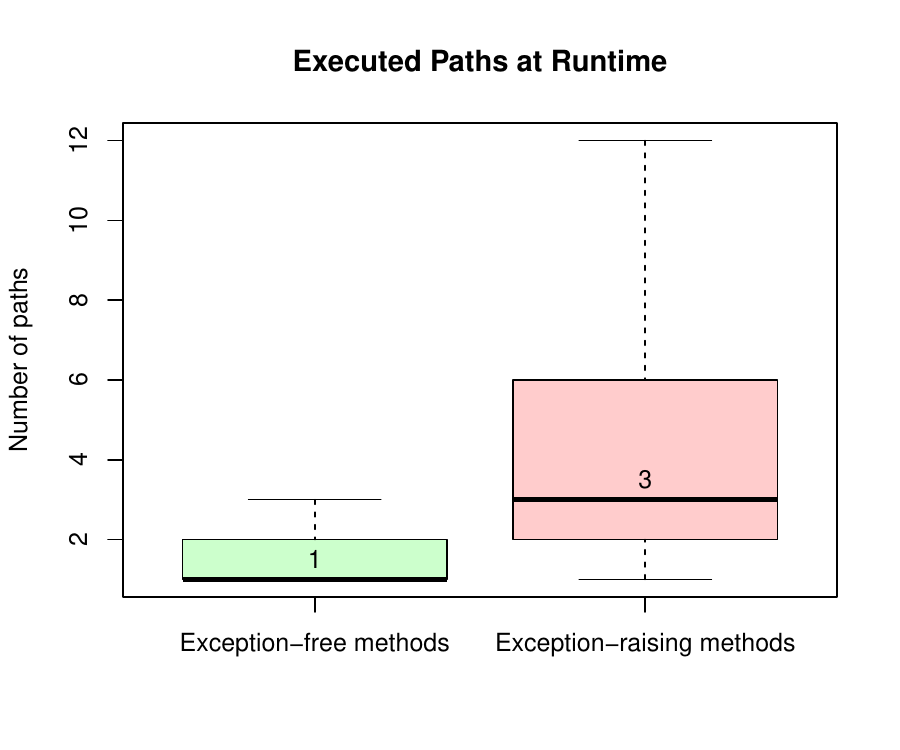}
    \caption{Distribution of executed paths at runtime.}
    \label{fig:box_paths}
\end{figure}

Figure~\ref{fig:box_calls} presents the distribution of method calls in exception-free and exception-raising methods.
On the median, the methods that do not raise exceptions have 24 calls, while those that do raise have 105 calls.
The difference is statistically significant (Mann-Whitney test, \emph{p-value $<$ 0.05}).
Figure~\ref{fig:box_paths} details the distribution of executed paths in both groups.
On the median, the methods that do not raise exceptions execute 1 path, while those that do raise execute 3 paths.
The difference is statistically significant (\emph{p-value $<$ 0.05}).

\begin{boxH}
\textbf{Observation 1}:
21.4\% of the executed methods raise exceptions at runtime.
On the median, exception-raising methods receive 4x more calls and execute 3x more paths than exception-free methods
\end{boxH}

\subsection{RQ2: How frequently do calls on exception-raising methods actually lead to exceptions?}

In this research question, we analyze the 1,150 methods that raise exceptions at runtime.
In the previous RQ, we saw that these methods received a median of 105 calls.
However, naturally, not necessarily all these calls raise exceptions at runtime.
Therefore, here, we focus on the method calls that actually lead to an exception being raised.

\begin{figure}[t]
    \centering
    \includegraphics[width=0.24\textwidth]{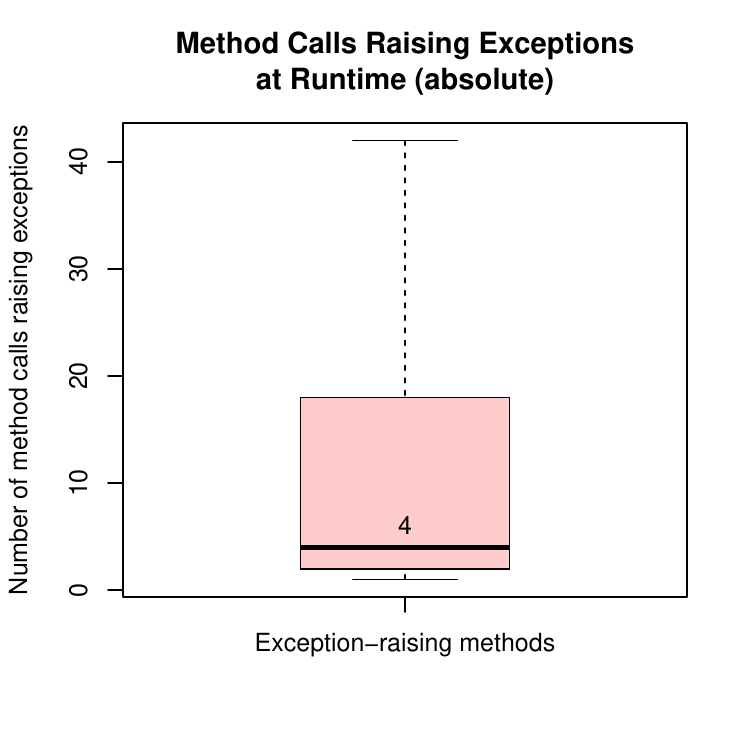}
    \includegraphics[width=0.24\textwidth]{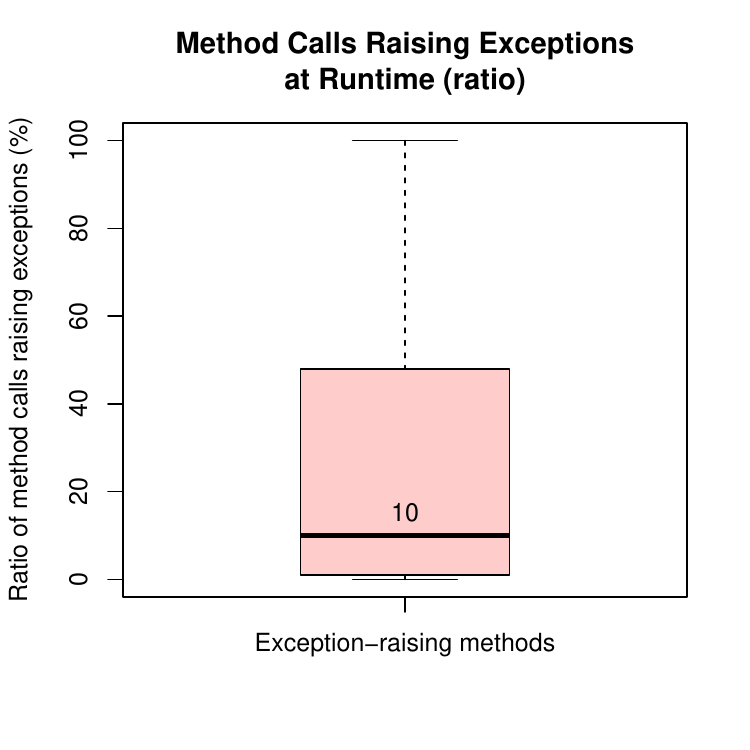}
    \caption{Distribution of the method calls raising exceptions at runtime (absolute and relative values).}
    \label{fig:box_calls_exe}
\end{figure}

Figure~\ref{fig:box_calls_exe} presents the distribution of method calls raising exceptions at runtime in both absolute (left-side) and relative (right-side) values.
On the median, exception-raising methods receive 4 calls that raise exceptions (the first quartile is 2 calls, and the third quartile is 18 calls).
In such methods, on the median, 10\% of the calls are exception-raising.
This means that 1 in 10 calls cover exceptional behaviors.
In this case, the first quartile is 1\%, while the third quartile is 48\%.
The first quartile in 1\% states that in 25\% of the exception-raising methods, at most 1 in 100 calls raise an exception.

\begin{boxH}
\textbf{Observation 2}:
In methods that raise exceptions, on the median, 1 in 10 calls exercise exceptional behaviors.
Each exception-raising method receives a median of 4 calls that cover exceptions.
\end{boxH}


Table~\ref{tab:frequency} groups the exception-raising methods into four categories according to their frequency of exception-raising calls: rare, occasional, common, and almost always.
Methods that raise exceptions in at most 10\% of their calls are categorized as \emph{rarely} raising exceptions.
Most exception-raising methods fall into this category, which includes 50\% (576) of the analyzed methods.
Methods that raise exceptions in more than 10\% and up to 50\% of their calls are classified as \emph{occasionally} raising exceptions.
We find that 28.4\% (327) of the methods raise exceptions occasionally.
Methods that raise exceptions in more than 50\% but less than 90\% of their calls are categorized as \emph{commonly} raising exceptions, while methods that raise exceptions in 90\% or more are classified as \emph{almost always} raising exceptions.
Only 9.6\% (111) of the methods commonly raise exceptions, while 11.8\% (136) almost always raise exceptions.

\begin{table}[t]
\centering
\caption{Frequency of exception-raising calls.}
\begin{tabular}{llcc}
\toprule
\textbf{Frequency} & \textbf{Exception-Raising Calls} & \textbf{\#Methods} & \textbf{\%} \\ \midrule
Rare               & $\le10\%$              & 576 & 50.0\% \\
Occasional         & $>10\%$ and $\le50\%$  & 327 & 28.4\% \\
Common           & $>50\%$ and $<90\%$    & 111 & 9.6\% \\
Almost always      & $\ge90\%$              & 136 & 11.8\% \\ \midrule
All & - & 1,150 & 100\% \\
\bottomrule
\end{tabular}
\label{tab:frequency}
\end{table}

\begin{boxH}
\textbf{Observation 3}:
Close to 80\% of the methods that raise exceptions at runtime do so infrequently, while
only about 20\% raise exceptions more frequently.
\end{boxH}

Table~\ref{tab:methods} presents multiple examples of exception-raising methods and their calls in each category.
Next, we briefly discuss some interesting examples.

\begin{table*}[t]
\centering
\caption{Examples of methods by frequency of exception-raising calls.}
\begin{tabular}{lll rrr}
\toprule
\multirow{2}{*}{\textbf{Frequency}} & \multirow{2}{*}{\textbf{Method}} & \multirow{2}{*}{\textbf{System}} & \multirow{2}{*}{\textbf{\#Calls}} & \multicolumn{2}{c}{\textbf{Exception-Raising Calls}} \\ \cmidrule{5-6}
&&&& \# & \% \\

\midrule

\multirow{10}{*}{Rare} & pylint.checkers.base\_checker.BaseChecker.create\_message\_definition\_from\_tuple	& pylint&	1,279,848 &	1 & $<$1\% \\
&email.\_header\_value\_parser.get\_ttext & email & 9,188 & 1 &	$<$1\% \\
&tarfile.TarInfo.\_decode\_pax\_field & tarfile & 25,092 & 194	& 1\% \\
&requests.utils.\_validate\_header\_part	& requests&	1,838 &	15 &	1\% \\
&dateutil.parser.\_parser.parser.parse&	dateutil	& 1,160	&36	& 3\% \\
&rich.live.Live.start	&rich	&26	&1&	4\% \\
&rich.console.Console.rule&	rich&	20&	1&	5\% \\
&flask.templating.render\_template	&flask&	32&	2&	6\% \\ 
&requests.sessions.Session.get	&requests&	34&	3&	9\% \\
&cookiecutter.repository.determine\_repo\_dir	&cookiecutter&	50&	5&	10\% \\

\midrule

\multirow{10}{*}{Occasional} &	cookiecutter.generate.apply\_overwrites\_to\_context	&	cookiecutter	&	19	&	2	&	11\% \\
&	flask.scaffold.Scaffold.get	&	flask	&	9	&	1	&	11\%	\\
&	pathlib.Path.touch	&	pathlib	&	17	&	2	&	12\%	\\
&	flask.app.Flask.url\_for	&	flask	&	42	&	7	&	17\%	\\
&	collections.namedtuple	&	collections	&	55	&	11	&	20\%	\\
&	logging.config.\_install\_handlers	&	logging	&	17	&	4	&	24\%	\\
&	requests.utils.is\_ipv4\_address	&	requests	&	52	&	15	&	29\%	\\
&	argparse.ArgumentParser.parse\_known\_args	&	argparse	&	4,432	&	1,347	&	30\%	\\
&	pylint.checkers.classes.class\_checker.ClassChecker.visit\_functiondef	&	pylint	&	3,521	&	1,247	&	35\%	\\
&	pathlib.PurePath.relative\_to	&	pathlib	&	303	&	140	&	46\%	\\

\midrule

\multirow{10}{*}{Common} &	pathlib.PurePath.\_\_hash\_\_	&	pathlib	&	12,414	&	6,384	&	51\%	\\
&	cookiecutter.zipfile.unzip	&	cookiecutter	&	16	&	9	&	56\%	\\
&	pylint.testutils.\_run.\_add\_rcfile\_default\_pylintrc	&	pylint	&	425	&	246	&	58\%	\\
&	configparser.RawConfigParser.\_get\_conv	&	configparser	&	1,274	&	765	&	60\%	\\
&	pylint.checkers.stdlib.StdlibChecker.\_check\_open\_call	&	pylint	&	135	&	84	&	62\%	\\
&	pathlib.Path.mkdir	&	pathlib	&	490	&	316	&	64\%	\\
&	pathlib.PurePath.with\_suffix	&	pathlib	&	91	&	63	&	69\%	\\
&	cookiecutter.replay.load	&	cookiecutter	&	4	&	3	&	75\%	\\
&	pathlib.Path.samefile	&	pathlib	&	20	&	16	&	80\%	\\
&	dateutil.parser.\_parser.parser.\_parse	&	dateutil	&	1,160	&	1,013	&	87\%	\\

\midrule

 &	pathlib.Path.rglob	&	pathlib	&	219	&	200	&	91\%	\\
&	email.\_header\_value\_parser.get\_address	& email &	334 &	306 &	92\% \\
&	pylint.utils.file\_state.FileState.set\_msg\_status	&	pylint	&	3,564	&	3,384	&	95\%	\\
&	requests.utils.get\_auth\_from\_url	&	requests	&	273	&	264	&	97\%	\\
Almost &	six.MovedModule.\_\_getattr\_\_	&	six	&	41	&	40	&	98\%	\\
Always &	email.message.Message.\_get\_params\_preserve	&	email	&	958	&	945	&	99\%	\\
 &	pathlib.\_RecursiveWildcardSelector.\_iterate\_directories	&	pathlib	&	6,085	&	6,025	&	99\%	\\
&	argparse.\_ActionsContainer.\_handle\_conflict\_error	&	argparse	&	4	&	4	&	100\%	\\
&	email.header.\_ValueFormatter.\_maxlengths	&	email	&	56	&	56	&	100\%	\\
&	pylint.utils.utils.get\_module\_and\_frameid	&	pylint	&	3,618	&	3,618	&	100\%	\\

\bottomrule
\end{tabular}
\label{tab:methods}
\end{table*}

\begin{figure}[h]
    \centering
    \includegraphics[width=0.48\textwidth]{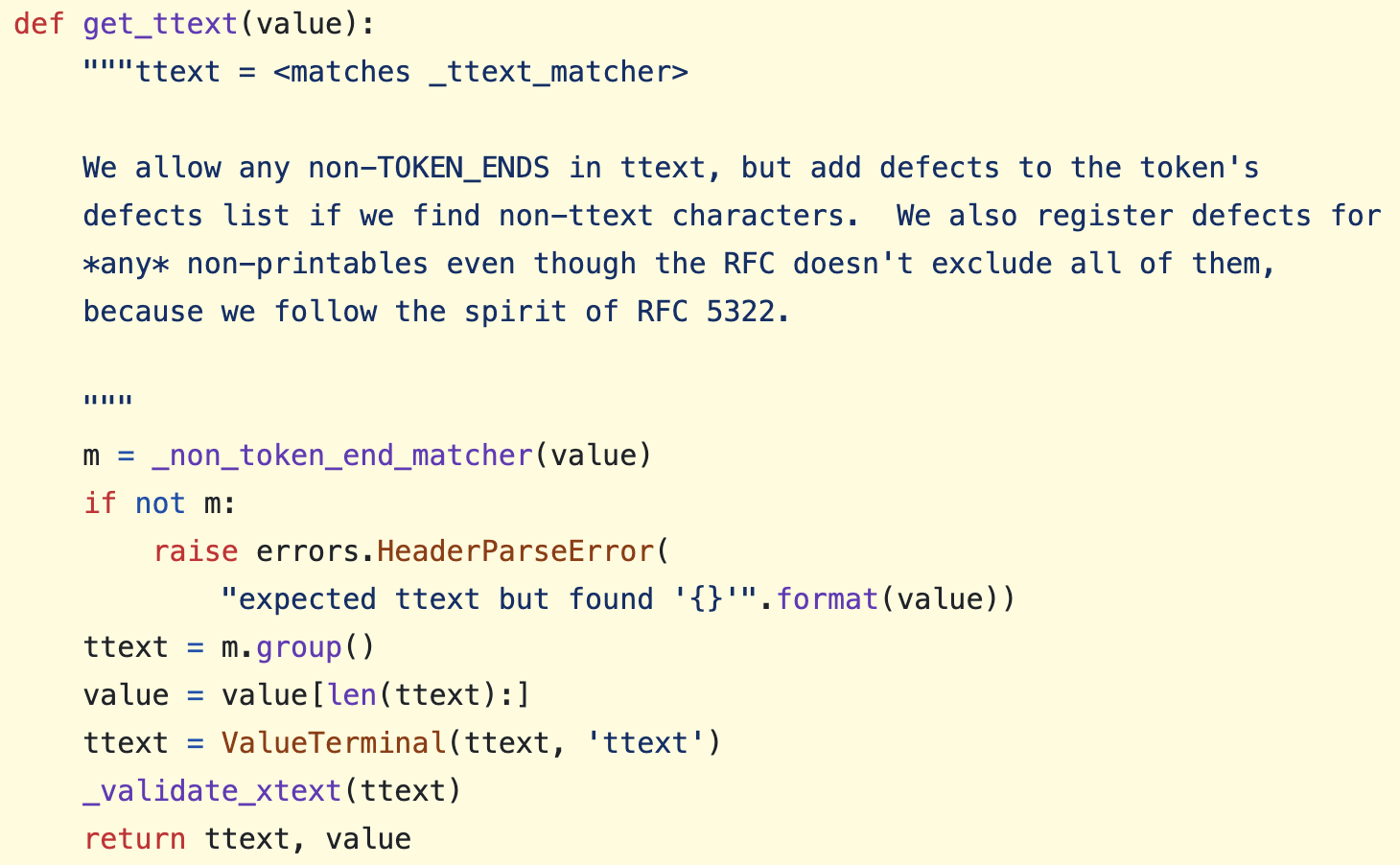}
    \caption{Method get\_ttext of the email library in CPython. Calls: 9,188; exception-raising calls: 1 ($<$1\%).}
    \label{fig:get_ttext}
\end{figure}

\textbf{Rarely raising exceptions:} 
This category includes methods that raise exceptions in at most 10\% of their calls.
The rarest exception found in our dataset happens in the Pylint method \texttt{create\_\-message\_\-definition\_\-from\_\-tuple}.\footnote{\url{https://github.com/pylint-dev/pylint/blob/c25923f3/pylint/checkers/base_checker.py\#L182}}
This method receives over 1,2 million calls, but only one call raises the \texttt{Invalid\-Message\-Error} due to a malformed message.
Method \texttt{get\_ttext}\footnote{\url{https://github.com/python/cpython/blob/e6dd71da/Lib/email/_header_value_parser.py\#L2258}} of the email library in CPython is another interesting example (as presented in Figure~\ref{fig:get_ttext}).
This method receives 9,188 calls, but only one call raises the exception \texttt{Header\-ParseError} due to an empty string passed to the parameter \texttt{value}.
Overall, this category includes methods in which raising the exception is a rare phenomenon.

\begin{figure}[h]
    \centering
    \includegraphics[width=0.45\textwidth]{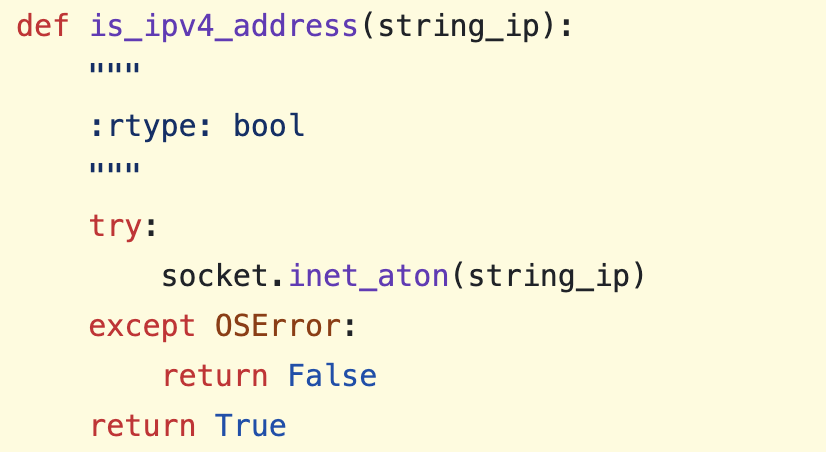}
    \caption{Method is\_ipv4\_address of Requests. Calls: 52; exception-raising calls: 15 (29\%).}
    \label{fig:is_ipv4_address}
\end{figure}

\textbf{Occasionally raising exceptions:} 
This category includes methods that raise exceptions in more than 10\% and up to 50\% of their calls.
As an example, we present the Requests method \texttt{is\_ipv4\_address}\footnote{\url{https://github.com/psf/requests/blob/7335bbf4/src/requests/utils.py\#L711}} (Figure~\ref{fig:is_ipv4_address}).
This method receives 52 calls, from which 15 (29\%) raise the exception \texttt{OSError}, indicating that the parameter \texttt{string\_ip} is an invalid IPv4.
Differently from the previous two examples (in which a single and rare call covered the exception), here, we see multiple exception-raising calls.
In this case, the 15 raised exceptions happened due to the following inputs: \emph{``localhost.localdomain''} (7 inputs), \emph{``www.requests.com''} (4 inputs), \emph{``google.com''} (3 inputs), and \emph{``8.8.8.8.8''} (1 input).
It is interesting to note that 2 out of the 15 exception-raising calls come directly from test \texttt{TestIsIPv4\-Address.\-test\_\-invalid}.\footnote{\url{https://github.com/psf/requests/blob/7335bbf4/tests/test_utils.py\#L261}}
In contrast, the other 13 exception-raising calls come indirectly from other tests, making them harder to track from the test perspective.

\begin{figure}[h]
    \centering
    \includegraphics[width=0.48\textwidth]{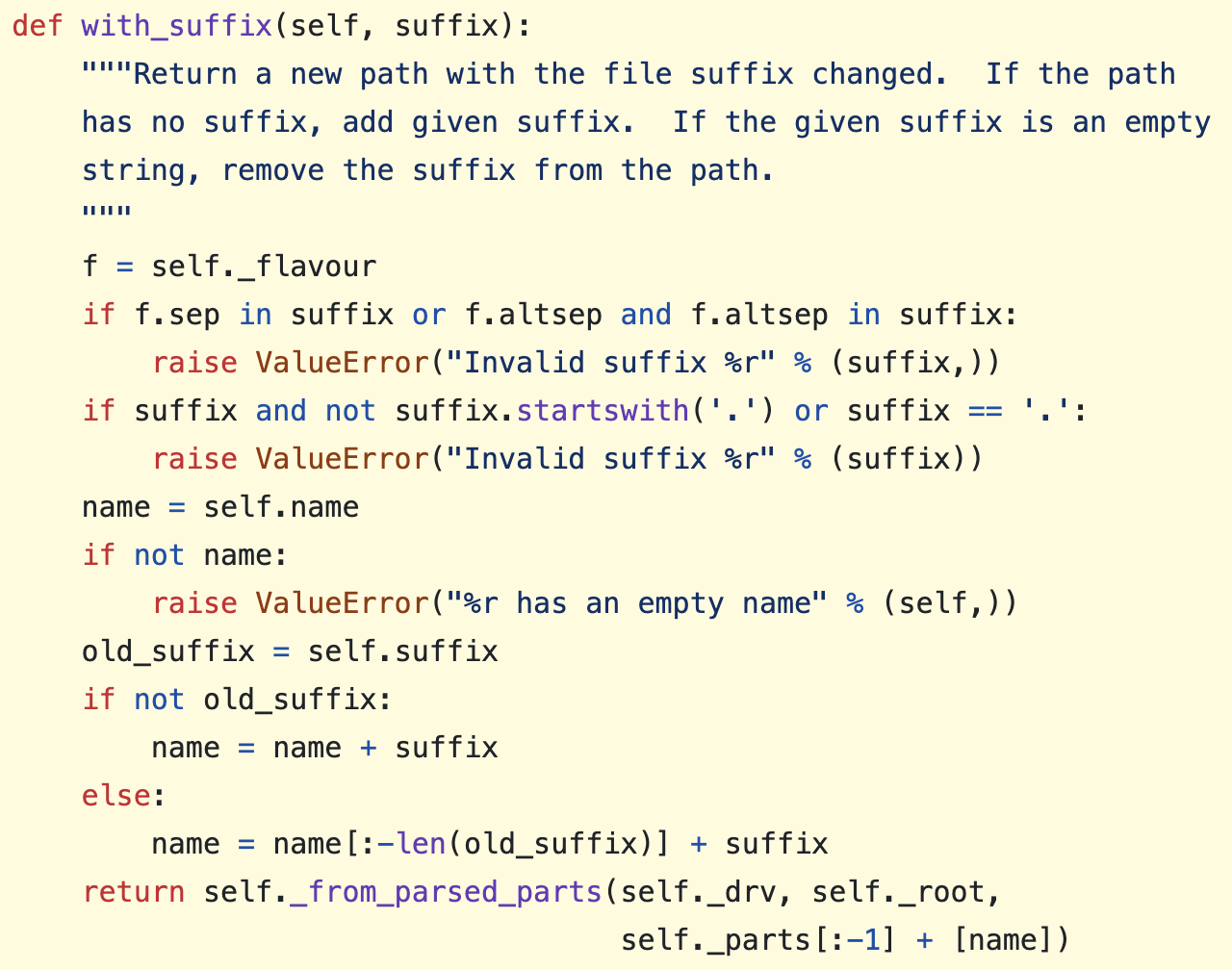}
    \caption{Method with\_suffix of the pathlib library in CPython. Calls: 91; exception-raising calls: 63 (69\%).}
    \label{fig:with_suffix}
\end{figure}

\textbf{Commonly raising exceptions:} 
This category includes methods that raise exceptions in more than 50\% but less than 90\% of their calls.
Method \texttt{with\_suffix}\footnote{\url{https://github.com/python/cpython/blob/0c5fc272/Lib/pathlib.py\#L770}} of the pathlib library in CPython presents an example in this category (Figure~\ref{fig:with_suffix}).
This method returns a new path with the file suffix changed, performing several validations in the parameter \texttt{suffix}.
Method \texttt{with\_suffix} receives 91 calls, from which 63 (69\%) raise the exception \texttt{ValueError}, indicating a problem in the suffix.
Unlike the previous two categories, here, raising the exception is more common than not raising it.
In fact, the presented method in Figure~\ref{fig:with_suffix} is more ``defensive'' in its implementation, with three \texttt{raise} statements.

\begin{figure}[h]
    \centering
    \includegraphics[width=0.48\textwidth]{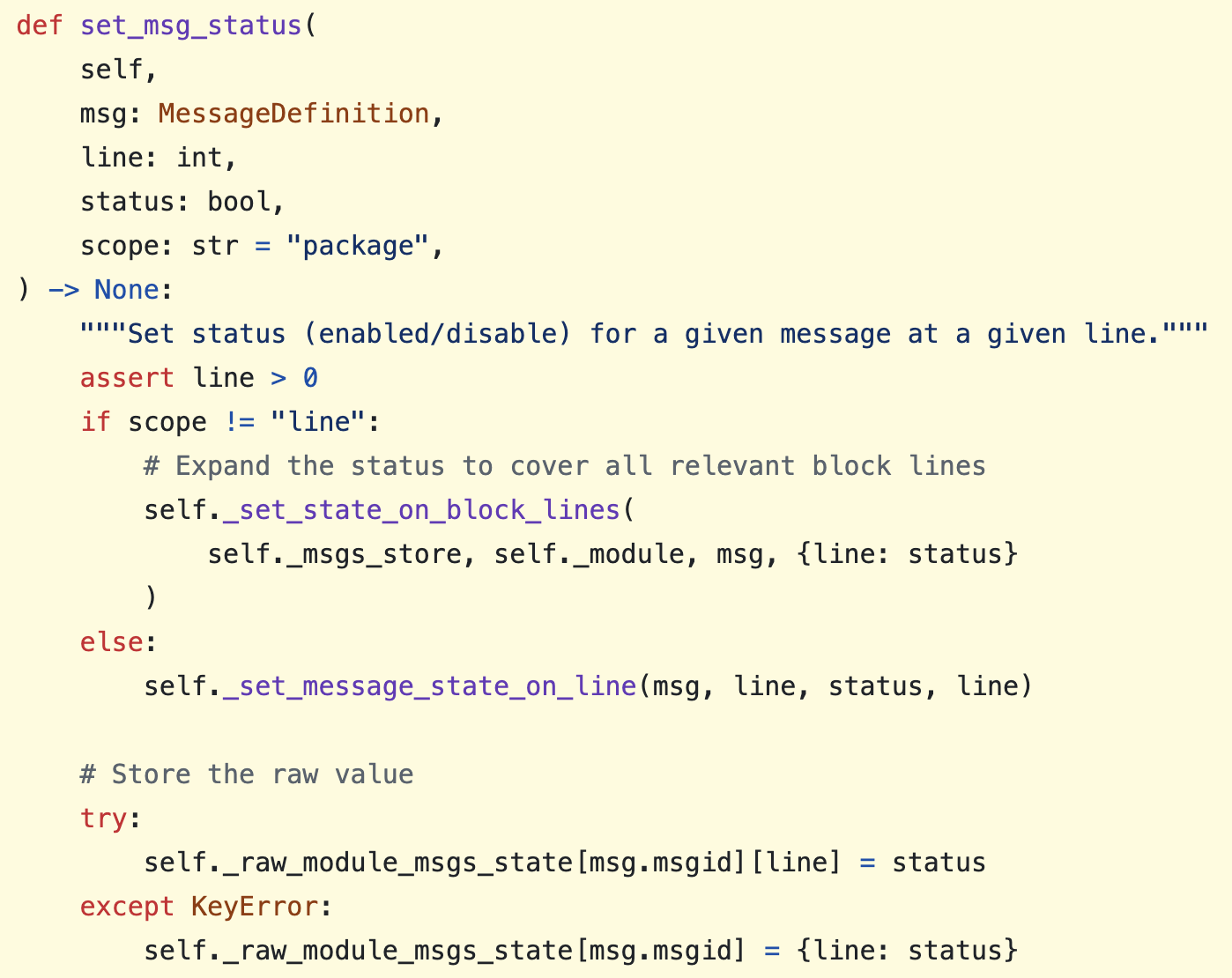}
    \caption{Method set\_msg\_status of Pylint. Calls: 3,564; exception-raising calls: 3,384 (95\%).}
    \label{fig:set_msg_status}
\end{figure}

\begin{figure}[h]
    \centering
    \includegraphics[width=0.48\textwidth]{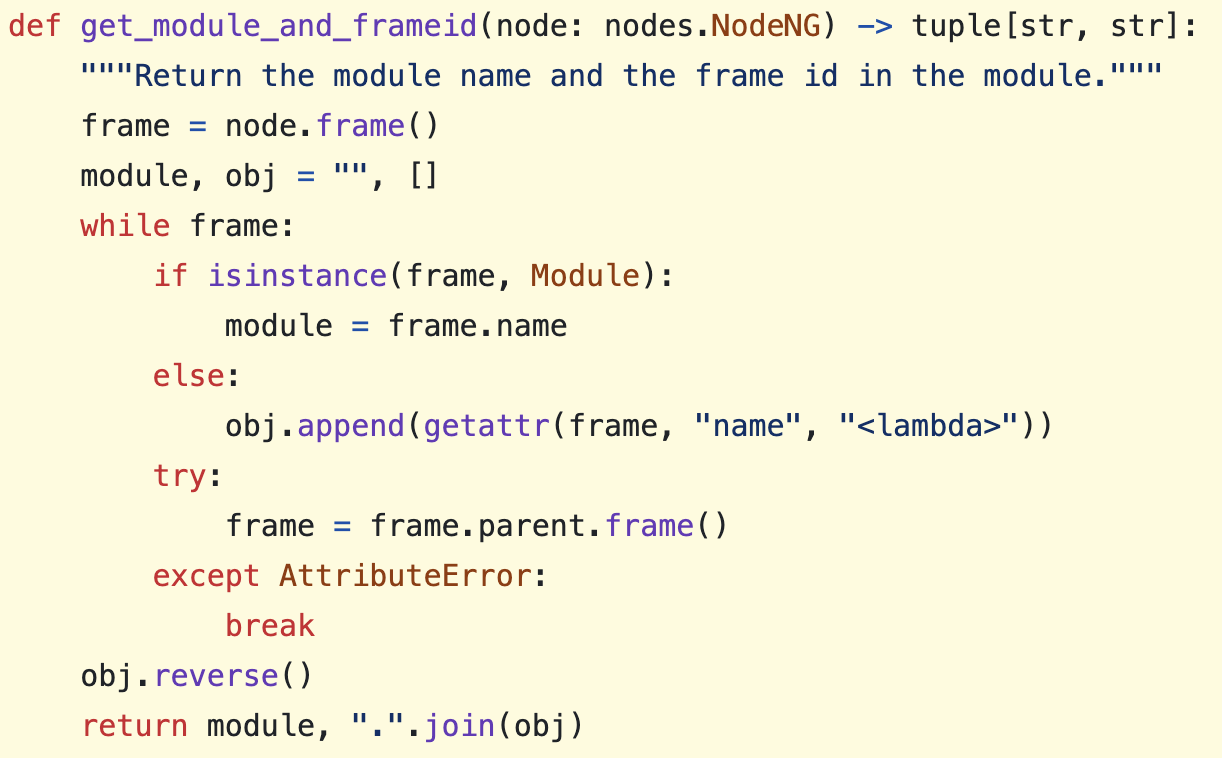}
    \caption{Method get\_module\_and\_frameid of Pylint. Calls: 3,618; exception-raising calls: 3,618 (100\%).}
    \label{fig:get_module_and_frameid}
\end{figure}

\textbf{Almost always raising exceptions:} 
This category includes methods that raise exceptions in 90\% or more of their calls.
We present two Pylint methods: \texttt{set\_\-msg\_\-status}\footnote{\url{https://github.com/pylint-dev/pylint/blob/c25923f3/pylint/utils/file_state.py\#L184}} and \texttt{get\_\-module\_\-and\_\-frameid}.\footnote{\url{https://github.com/pylint-dev/pylint/blob/c25923f3/pylint/utils/utils.py\#L103}}
In method \texttt{set\_msg\_status} (Figure~\ref{fig:set_msg_status}), 3,384 in 3,564 (95\%) calls raise the exception \texttt{KeyError}.
Figure~\ref{fig:get_module_and_frameid} presents an even more strict case: 100\% (all 3,618) of the calls to method \texttt{get\_\-module\_\-and\_\-frameid} raise the exception \texttt{Attribute\-Error}.
In this case, raising the exception is a terminating condition for the \texttt{while} loop.
It is interesting to note that in such methods, the ``normal'' behavior is actually to raise the exception.
That is, the methods are implemented in such a way that raising is part of the expected behavior.

\begin{boxH}
\textbf{Observation 4}:
Typically, exception-raising at runtime is a rare phenomenon and indicates ``abnormal'' behaviors.
However, in some cases, exception-raising is frequent and represents ``normal'' behaviors. 
\end{boxH}

\subsection{RQ3: How do exception-raising methods and calls vary by system?}

In this last RQ, we explore the variation of exception-raising methods, and calls vary per system.
Table~\ref{tab:per_system} details the exception-raising methods and calls per system.
Column \emph{``Exception-Raising Methods''} presents the absolute (\#) and relative (\%) number of exception-raising methods.
Column \emph{``Exception-Raising Calls''} presents the distribution of exception-raising calls for the exception-raising methods, detailing the first quartile (Q1), second quartile (Q2), and third quartile (Q3).
For example, Table~\ref{tab:per_system} shows that Pylint's tests execute 1,537 methods, of which 273 (17.8\%) are exception-raising.
Among these 273 exception-raising methods, the median proportion of exception-raising calls is 7\%, with a first quartile of 1\% and a third quartile of 50\%.

\begin{table*}[t]
\centering
\caption{Exception-raising methods and calls by system. Q1: first quartile; Q2: second quartile (median); Q3: third quartile}
\begin{tabular}{l rrr c rrr}
\toprule
\multirow{2}{*}{\textbf{System}} & \textbf{Executed} & \multicolumn{2}{c}{\textbf{Exception-Raising Methods}} && \multicolumn{3}{c}{\textbf{Exception-Raising Calls (\%)}} \\ \cmidrule{3-4} \cmidrule{6-8}
& \textbf{Methods} & \textbf{\#} & \textbf{\%} && \textbf{Q1} & \textbf{Q2} & \textbf{Q3} \\  \midrule
Pylint & 1,537 & 273 & 17.8\% && 1\% & 7\% & 50\% \\ 
Rich & 589 & 73 & 12.4\% && 1\% & 7\% & 29\% \\ 
Error Corrector & 550 & 31 & 5.6\% && 18.5\% & 50\% & 80\% \\ 
BentoML & 319 & 56 & 17.6\% && 8\% & 27\% & 72\% \\ 
Flask & 284 & 78 & 27.5\% && 4\% & 15\% & 70\% \\ 
Dateutil & 241 & 63 & 26.1\% && 2\% & 9\% & 22\% \\ 
Requests & 174 & 44 & 25.3\% && 2\% & 12.5\% & 34\% \\ 
Jupyter Client & 138 & 31 & 22.5\% && 21\% & 100\% & 100\% \\ 
Cookiecutter & 66 & 35 & 53\% && 6\% & 12\% & 31\% \\ 
Six & 32 & 7 & 21.9\% && 1\% & 60\% & 99\% \\ 
\midrule
email & 381 & 120 & 31.5\% && 1\% & 4\% & 25\% \\ 
logging & 216 & 49 & 22.7\% && $<$1\% & 5\% & 24\% \\ 
argparse & 126 & 38 & 30.2\% && $<$1\% & 2\% & 31\% \\ 
collections & 112 & 26 & 23.2\% && 9\% & 58\% & 100\% \\ 
pathlib & 97 & 42 & 43.3\% && 7\% & 41.5\% & 62\% \\ 
tarfile & 89 & 33 & 37.1\% && $<$1\% & 5\% & 19\% \\ 
configparser & 82 & 41 & 50\% && 2\% & 8\% & 23\% \\ 
calendar & 63 & 4 & 6.3\% && 1\% & 1\% & 11\% \\ 
ftplib & 51 & 21 & 41.2\% && 1\% & 4\% & 33\% \\ 
difflib & 47 & 9 & 19.1\% && 16\% & 22\% & 44\% \\ 
imaplib & 47 & 18 & 38.3\% && 3\% & 4\% & 18\% \\ 
smtplib & 43 & 19 & 44.2\% && 4\% & 8\% & 31\% \\ 
os & 41 & 20 & 48.8\% && 10\% & 45\% & 75\% \\ 
gzip & 32 & 11 & 34.4\% && 1\% & 2\% & 3\% \\ 
csv & 15 & 8 & 53.3\% && 2\% & 9\% & 71\% \\ \midrule
All & 5,372 & 1,150 & 21.4\% && 1\% & 10\% & 48\% \\ 
\bottomrule
\end{tabular}
\label{tab:per_system}
\end{table*}

\textbf{Exception-raising methods:}
Overall, we notice a large variation in the ratio of exception-raising methods per system, ranging from 5.6\% to 53.3\%.
The system with the smallest proportion of exception-raising methods is Error Corrector, with 5.6\% (31 out of 550 methods), while the system with the largest proportion is csv, with 53.3\% (8 out of 15 methods).

It is worth noting that only 3 out of 25 systems (Cookiecutter, configparser, and csv) have 50\% or more exception-raising methods.
In contrast, the remaining 22 systems have less than 50\% exception-raising methods.
Among those, two systems have less than 10\%. 

\begin{boxH}
\textbf{Observation 5}:
The majority of systems (22 out of 25) contain more exception-free methods than exception-raising methods.
\end{boxH}

\textbf{Exception-raising calls:}
We also find a large variation in the proportion of exception-raising calls.
The median of exception-raising calls per method varies from 1\% (calendar) to 100\% (Jupyter Client), the first quartile varies from $<$1\% (logging, argparse, and tarfile) to 21\% (Jupyter Client), and the third quartile varies from 3\% (gzip) to 100\% (Jupyter Client).
Among the 25 systems, 19 (76\%) have a median value of less than 30\%, while only 6 systems (24\%) have a median value greater than 30\%.
Notice that not all systems have an equivalent number of exception-raising methods.
For instance, Pylint has 273 exception-raising methods, while calendar has only 4.
Consider only the top-5 systems with the most exception-raising methods: Pylint (273), email (120), Flask (78), Rich (73), and Dateutil (63).
In this case, the median of exception-raising calls per method is much smaller, ranging from 4\% (email) to 15\% (Flask).

\begin{boxH}
\textbf{Observation 6}:
Most systems (19 out of 25) have a median proportion of exception-raising calls per method below 30\%.
\end{boxH}

\section{Discussion and Implications}

\subsection{Most Exceptional Behaviors Are Rarely Exercised}

The literature reports that developers are more likely to test expected behaviors than unexpected ones, such as exceptional behaviors~\cite{teasley1994software, salman2019controlled, causevic2013effects, edwards2014student, aniche2021developers, mohanani2018cognitive, leventhal1993positive, bijlsma2021students, garousi2018smells, bai2021students, goffi2016automatic}.
In this study, we contribute to this research line by showing that most methods that raise exceptions at runtime do so infrequently.
In other words, given a method that raises an exception at runtime under specific conditions, it is likely that such an exception will rarely be triggered by the existing test suite.
For example, we find that 50\% of the exception-raising methods actually raise exceptions in at most 10\% of their calls.
This can be problematic because the fact that error handling code such as exceptions are not often executed makes them a natural place to hide bugs~\cite{fix_error_handling_first, goffi2016automatic}.

\underline{\textbf{Implication 1}}:
We envision the development of novel tools to support exercising exceptional behaviors more effectively.
For example, such tools could identify the tests that cover exceptional cases, including exceptions not propagated to tests.
Such tools could alert developers whether exceptional cases are adequately tested or missing in the test suite.

\subsection{Some Exceptional Behaviors Are Frequently Exercised}

The majority of methods that raise exceptions at runtime do so infrequently.
However, some methods actually present the opposite behavior: they raise exceptions more frequently.
That is, given a method that raises an exception at runtime, in rare cases, the exception is frequently triggered by the existing test suite.
For example, we find that 11.8\% of the exception-raising methods raise exceptions in 90\% or more of their calls.
For such methods, the ``normal'' behavior is to raise the exception.

One factor that may explain such a variation is the origin of the exception.
For example, exceptions can be explicitly raised by the system under test (as in Figure~\ref{fig:ex1a}), or implicitly raised by external dependencies or standard libraries (as in Figure~\ref{fig:ex1b})~\cite{marcilio2021java, mcconnell2004code}.
An exception raised directly by the system seems more relevant and testable behavior, whereas an exception raised somewhere deeper in the call stack may indicate less direct relevance.

Although we did not deeply explore the origin of the exceptions, RQ1 provides initial insights in this direction.
Table~\ref{tab:exe_types} shows that most exceptions are raised by up to three methods, suggesting they are potentially specific, like the ones originating from the system under test (\eg~\texttt{SMTP\-Sender\-Refused}).
In contrast, a few exceptions are raised by ten or more methods, indicating they are generic, like the ones originating from standard libraries (\eg~\texttt{Value\-Error}).

\underline{\textbf{Implication 2}}:
We call attention to the fact that exception-raising behaviors are not necessarily ``abnormal'' or rare.
The fact that a test exercises a method that raises exceptions does not necessarily indicate it is testing an ``abnormal'' behavior.
Raising an exception may simply be part of the method's ``normal'' behavior.
Researchers working on exceptional behavior testing~\cite{dalton2020exceptional, marcilio2021java, lima2021assessing, zhang2024generating, yoshioka2023exceptional} should be aware of such methods to avoid failing to detect abnormal behaviors.
In such cases, it would be important to consider other factors, such as the origin of the raised exceptions (\ie~SUT or external).

\subsection{Refactoring Expensive \texttt{try/except} Blocks}

Our results show that different exception-raising methods may have distinct frequencies of exception-raising calls, ranging from rare to almost always.
One explanation may be the Python coding style EAFP (easier to ask for forgiveness than permission), which assumes the existence of valid keys or attributes, catching the exceptions \texttt{KeyError} or \texttt{AttributesError} if the assumption proves false.\footnote{\url{https://docs.python.org/3/glossary.html\#term-EAFP}}
According to the Python documentation, this coding style is characterized by the presence of many \texttt{try/except} blocks.
However, note that \texttt{try/except} blocks are very efficient when no exceptions are raised, but catching an exception is actually expensive.\footnote{\url{https://docs.python.org/3/faq/design.html\#how-fast-are-exceptions}}
Thus, a \texttt{try/except} block that checks the existence of a key should expect the dictionary to have the key almost all the time and rarely raise the exception.
This way, one possible refactoring is replacing such \texttt{try/except} blocks with more efficient solutions, \eg~checking the key existence with the \texttt{in} keyword in \texttt{if/else} blocks.

\begin{figure}[h]
    \centering
    \includegraphics[width=0.48\textwidth]{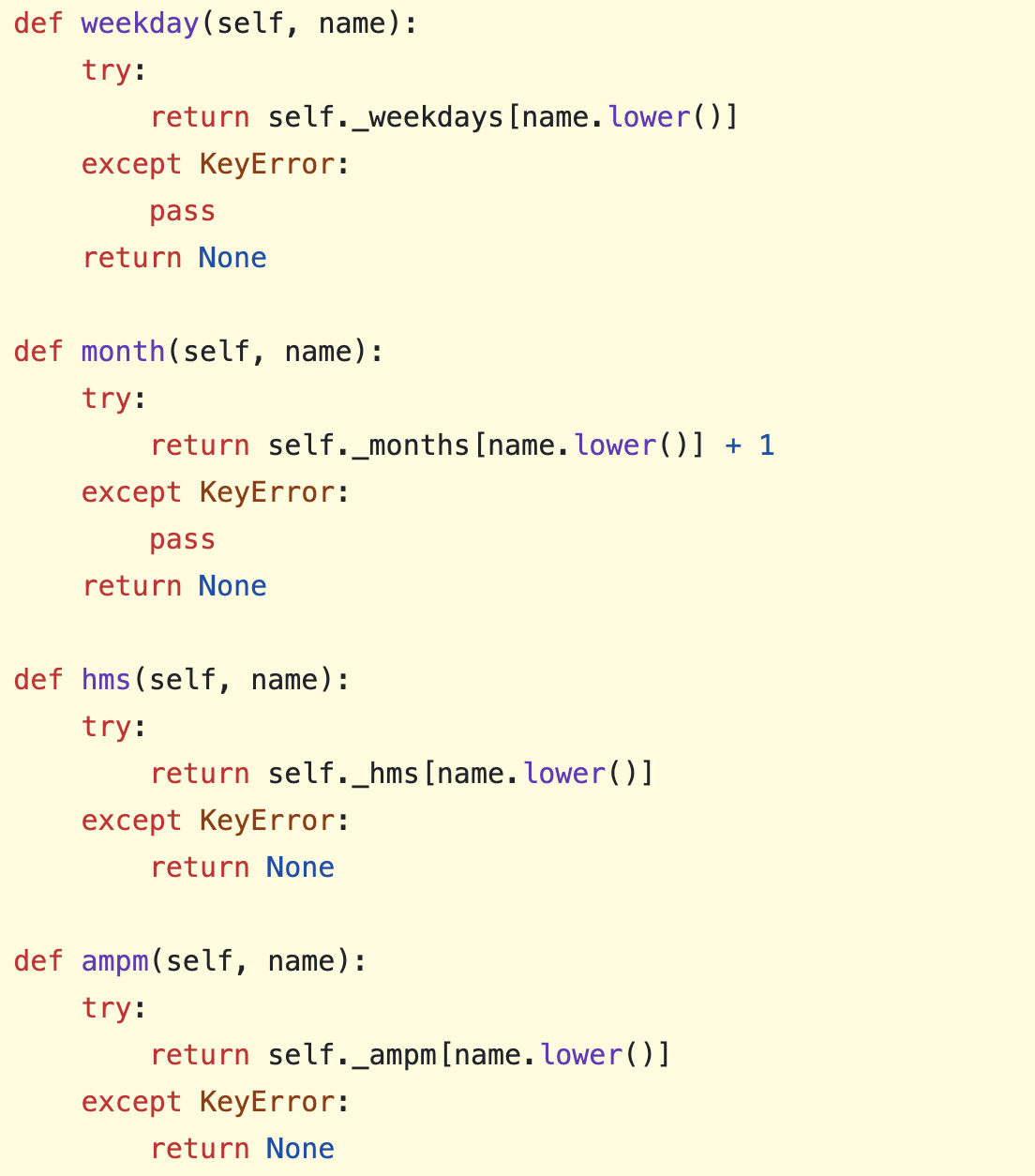}
    \caption{Methods that frequently raise the exception \texttt{Key\-Error}. The \texttt{try/except} blocks can be replaced by \texttt{if/else} blocks to increase efficiency.}
    \label{fig:_parser}
\end{figure}

Interestingly, we find that many methods commonly raise the exceptions \texttt{KeyError} and \texttt{AttributesError}, indicating they are candidates to be refactored.
For example, Figure~\ref{fig:_parser} shows four methods of project DateUtil that frequently raise \texttt{KeyError}: \texttt{ampm} (it raises exceptions in 96\% of the calls), \texttt{hms} (96\%), \texttt{weekday} (83\%), and \texttt{month} (75\%).\footnote{\url{https://github.com/dateutil/dateutil/blob/9eaa5de584f9f374/src/dateutil/parser/_parser.py\#L322-L346}}
Another extreme case happens in method \texttt{\_infer\_dunder\_doc\_attribute}\footnote{\url{https://github.com/pylint-dev/pylint/blob/5c59b48acb5e0c8e/pylint/checkers/base/docstring_checker.py\#L32-L35}} of project Pylint.
In this method, the exception \texttt{KeyError} happens in 99.9\% (2,729 of 2,731) of the calls. 
Therefore, such methods could be refactored to check the existence of the key using the \texttt{in} keyword and \texttt{if/else} blocks to increase efficiency.

\underline{\textbf{Implication 3}}:
\texttt{try/except} blocks that frequently raise exceptions at runtime when checking the existence of keys or attributes are strong candidates for refactoring.
Our study can spot those less efficient \texttt{try/except} blocks that can be replaced by more efficient \texttt{if/else} blocks checking the key existence.
In this context, we envision that future research in the context of refactoring~\cite{tsantalis2009identification, al2015identifying, alomar2024refactor, fan2023large, pomian2024together, shirafuji2023refactoring} could leverage the execution frequency of language constructors (\eg~\texttt{try/except} blocks) to detect novel refactoring opportunities.

\section{Threats to Validity}

\noindent\textit{Origin of the exceptions}. 
One factor that may affect the variation of the exception-raising methods and calls is their origin.
For example, exceptions directly raised by the SUT seem more relevant and testable than those raised by third-party libraries and propagated to the SUT~\cite{marcilio2021java, mcconnell2004code}.
Although we provided initial insights into the type of exceptions, further studies are needed to better understand the effects of the origin of the exceptions.

\noindent\textit{Public APIs and implementation details}. 
In this study, we do not distinguish between public APIs (\eg~public methods) and implementation details (\eg~private or internal methods) when exploring the exceptions raised at runtime.
Implementation details exist to support public APIs~\cite{public_apis_google}.
While public APIs should be well-tested, implementation details can sometimes be indirectly tested by public APIs~\cite{public_apis_google}.
However, complex implementation details can be directly tested in test suites~\cite{public_apis_google}.
Further studies are needed to explore the differences between exception-raising in public APIs and implementation details.

\noindent\textit{Generalization of the results}. 
In this study, we analyzed exception-raising from 25 popular and real-world Python test suites.
Despite these observations, our findings -- as usual in empirical software engineering -- cannot be directly generalized to other systems or implemented in other programming languages.
Further studies should be performed on other software ecosystems and programming languages.

\section{Related Work}

Overall, the literature agrees that developers are more likely to test normal behaviors than abnormal ones~\cite{teasley1994software, salman2019controlled, causevic2013effects, edwards2014student, aniche2021developers, mohanani2018cognitive, leventhal1993positive, bijlsma2021students, garousi2018smells, bai2021students, goffi2016automatic, hora2024monitoring}.
This happens because the expected behavior of the program is often simpler to test.
Another factor is that developers may lack test expertise, focusing on only testing the ``happy cases''~\cite{edwards2014student}.
In an experiment with developers, Teasley~\etal~\cite{teasley1994software} found evidence of using a positive test strategy (\ie~testing the expected behavior), which was partially mitigated by increasing the expertise of the developers.

Many studies explore exceptional behaviors from the test perceptive~\cite{dalton2020exceptional, marcilio2021java, lima2021assessing, zhang2024generating, yoshioka2023exceptional, goffi2016automatic, artho2006enforcer}.
Goffi~\etal~created test oracles for exceptional behaviors from Javadoc comments~\cite{goffi2016automatic}.
The authors also reported that exceptional behaviors are poorly covered by tests.
Lima~\etal~explored exception handling testing practices in Java libraries and found that \texttt{catch} blocks are less covered~\cite{lima2021assessing}.
Marcilio and Furia provided a large-scale empirical study of exceptional tests in Java, that is, tests that may trigger exceptional behaviors~\cite{marcilio2021java}.
The authors analyze multiple patterns Java developers can use to write exceptional tests and detect several characteristics of such tests.
For example, exceptional tests represent 13\% of all tests, tend to be larger, and are mostly written using \texttt{try/catch} blocks.
In this context, Dalton~\etal~\cite{dalton2020exceptional} analyzed exceptional behavior testing in Java, that is, tests that expect exceptions to be raised.
The results showed that 60.9\% of the projects have at least one test dedicated to verifying exceptional behavior, concluding that exceptional behavior testing is a rare phenomenon.
In common, both studies~\cite{marcilio2021java, dalton2020exceptional} explore the tests that expect exceptions to be raised.

Despite the various studies on exceptional behavior testing, they are mainly concentrated on analyzing raised exceptions that propagated to tests.
Our study provides a complementary perspective: we analyzed all raised exceptions at runtime, not only the ones that propagate to tests.
Furthermore, we deeply explored the frequency of raised exceptions at runtime, which is not the focus of any prior study.




\section{Conclusion}

We provided an empirical study to explore how frequently exceptional behaviors are tested in real-world systems.
We analyzed the test suites of 25 Python systems, covering 5,372 executed methods, 17.9M calls, and 1.4M raised exceptions.
Our main findings can be summarized as follows:
(1) 21.4\% of the executed methods do raise exceptions at runtime;
(2) in methods that raise exceptions, on the median, 1 in 10 calls exercise the exceptional behaviors; and
(3) close to 80\% of the methods that raise exceptions at runtime do so infrequently, while only about 20\% raise exceptions more frequently; and
(4) most systems contain more exception-free methods than exception-raising methods.
Based on our results, we discussed practical implications for researchers and practitioners, including the development of novel tools to more effectively support exercising exceptional behaviors and the refactoring of expensive \texttt{try/except} blocks.

In future work, we plan to perform more qualitative analysis on the analyzed methods, calls, and exceptions, for example, by exploring the origin of the raised exceptions (\ie~SUT or external)~\cite{marcilio2021java, mcconnell2004code} and whether they come from public APIs or implementation details~\cite{public_apis_google}.
We also intend to develop tools to identify the tests that cover exceptional cases, including exceptions not propagated to tests.
Finally, we plan to provide an empirical study to quantify and qualify the refactoring to replace expensive \texttt{try/except} blocks.

\section*{Acknowledgments}

This research is supported by CAPES, CNPq, and FAPEMIG.

\bibliographystyle{IEEEtran}
\bibliography{main}

\end{document}